\documentclass{ws-ijmpd}

\begin{document}

\markboth{A. Pulido Pat\'{o}n} {ASTROD Inertial Sensor}

%%%%%%%%%%%%%%%%%%%%% Publisher's Area please ignore %%%%%%%%%%%%%%%
%
\catchline{}{}{}{}{}
%
%%%%%%%%%%%%%%%%%%%%%%%%%%%%%%%%%%%%%%%%%%%%%%%%%%%%%%%%%%%%%%%%%%%%

\title{Current prospects for ASTROD Inertial Sensor}

\author{A. PULIDO PAT\'{O}N}

\address{Center for Gravitation and Cosmology, Purple Mountain Observatory,\\ Chinese Academy of Sciences, Beijing West Road 2,\\
Nanjing 210008, China.\\
antonio@pmo.ac.cn}

\maketitle

\begin{abstract}
The Astrodynamical Space Test of Relativity using Optical Devices
(ASTROD) is a multi-purpose relativity mission concept. ASTROD's
scientific goals are the measurement of relativistic and solar
system parameters to unprecedented precision, and the detection
and observation of low-frequency gravitational waves to
frequencies down to $5\times10^{-6}$ Hz. To accomplish its goals,
ASTROD will employ a constellation of drag-free satellites, aiming
for a residual acceleration noise of (0.3-1)$\times$ 10$^{-15}$ m
s$^{-2}$ Hz$^{-1/2}$ at 0.1 mHz. Noise sources and strategies for
improving present acceleration noise levels are reported.

\end{abstract}

\keywords{ASTROD; LISA; Space interferometers; Inertial sensors.}

\section{Introduction}

The classical concept of a drag-free satellite\cite{Lange}
consists of a small proof mass inserted inside a large spacecraft.
The larger spacecraft shields external forces, allowing the proof
mass to move in free fall. The relative position and orientation
of the proof mass with respect to the main spacecraft is sensed
along its trajectory. This information is feedback to thrusters on
the main spacecraft, which are subsequently fired to maintain the
proof mass-spacecraft relative position and orientation. In this
way, the coupling of external forces to the proof mass is
minimized.

Numerous space missions have already employed drag-free
technology. The first example of a "drag-free" mission to test
fundamental physics is Gravity Probe B (GP-B).\cite{GPB} GP-B was
launched in April 2004 and has experimentally measured
frame-dragging and geodetic effects. By spring of 2007, the
results of the data analysis will become public. The GP-B inertial
sensor achieved a level of free fall below $2\times10^{-12}$ m
s$^{-2}$ Hz$^{-1/2}$ at $5\times10^{-3}$ Hz.

The Laser Interferometer Space Antenna, LISA, is another
fundamental physics mission requiring drag-free performance. The
LISA mission concept\cite{LISA} consists of three spacecraft in
heliocentric orbits, forming a nearly equilateral triangle
formation of side 5 $\times$ 10$^{6}$ km and inclined with respect
to the ecliptic by 60$^{\circ}$. LISA will monitor the separation
between free falling proof masses, which are shielded within the
spacecraft, by using interferometric techniques, to detect and
observe gravitational waves. LISA's aims include studying the role
of massive black holes in galaxy evolution, testing relativistic
gravity, determining the population of ultra-compact galactic
binaries, probing the physics of the early universe, observing
supermassive and intermediate black holes mergers and mapping
spacetime by observing gravitational captures. The LISA drag-free
performance goal is 3 $\times$ 10$^{-15}$ m s$^{-2}$ Hz$^{-1/2}$
at 0.1 mHz. By 2009, the LISA Technology Package (LTP) on board
the LISA Pathfinder (LPF) ESA mission, with NASA contributions,
aims to demonstrate drag-free performance to a level one order of
magnitude lower than the LISA requirement, approximately 3
$\times$ 10$^{-14}$ m s$^{-2}$ Hz$^{-1/2}$ in the frequency
bandwidth between 1 mHz and 30 mHz.\cite{LTP} If LTP is
successful, it will achieve the best drag-free performance up to
present date.

The Astrodynamical Space Test of Relativity using Optical Devices,
ASTROD, is a mission concept\cite{ASTROD,Ni} that consists of a
constellation of drag-free spacecraft employing laser
interferometric techniques with Earth orbiting satellites, to
provide high precision measurements of the relativistic parameters
\{$\gamma$, $\beta$\}; improved determination of the orbits of
major asteroids; measurement of solar angular momentum via the
Lense-Thirring effect and the detection of low-frequency
gravitational waves and solar oscillations. ASTROD aims to improve
on the LISA drag-free goal by a factor of between 3 and 10, i.e.
(0.3-1) $\times$ 10$^{-15}$ m s$^{-2}$ Hz$^{-1/2}$ at 0.1
mHz.\cite{NiGRG,NiNPBProc,NiIJMPD07,NiLiao,Pulido1,Pulido2} It is
worth noting that a ten-fold acceleration noise improvement with
respect to LISA, would allow ASTROD to explore relativistic
gravity to an uncertainty level of 1
ppb.\cite{NiGRG,NiNPBProc,PPBASTROD,NiGravEmp} A simple version of
ASTROD, ASTROD I, has been studied as the first step to ASTROD.
ASTROD I concept consists of one spacecraft in a solar orbit,
carrying out interferometric ranging and pulse ranging with ground
stations.\cite{NiGRG,NiNPBProc,NiIJMPD07} ASTROD I also aims to
measure relativistic and solar system parameters and test
fundamental laws of spacetime although with less precision than
ASTROD.

Other mission concepts to follow on from LISA are BBO\cite{BBO}
(Big Bang Observer) and DECIGO\cite{DECIGO} (DECihertz
Interferometry Gravitational wave Observatory). These missions aim
to observe the cosmic gravitational wave background produced by
standard inflation, having optimum sensitivity around 0.1 Hz,
where white dwarf binaries confusion level is thought to be very
low. The drag-free force noise requirement for BBO and DECIGO is
approximately a hundredth of the LISA force noise target.

In section 2 we report general ideas that ASTROD could adopt to
improve drag-free performance, we summarize acceleration and
sensor back action disturbances, emphasizing the most significant
low-frequency noise sources. Charging disturbances and discharging
schemes are briefly discussed in section 3. Finally in section 4
we discuss the performance and problems associated with inertial
sensors for other follow on LISA missions. Gravitational
interactions are described in terms of multipole moments and a
detailed calculation of capacitances for the case of a three
dimensional capacitive sensor/actuator are given in appendix A and
B, respectively.

\section{ASTROD Inertial Sensor}

ASTROD will face new challenges in  drag-free technology compared
with LISA. First, ASTROD aims to improve the LISA drag-free
performance target by a factor of between 3 and 10 at 0.1 mHz. On
the other hand, ASTROD will extend the gravitational wave
observational bandwidth to frequencies below 0.1 mHz, which is the
lowest frequency on the LISA observational bandwidth, down to 5
$\times$ 10$^{-6}$ Hz. To achieve these goals, there are key
problems that need to be evaluated and their solutions optimized.

The initial design concept proposed for LISA was that laser beams
from remote spacecraft would directly illuminate the proof masses.
Issues like laser beam pointing, actuation forces to correctly
orientate the proof mass with respect to laser beam, and cross
coupling of proof mass degrees of freedom, make this scheme less
atractive. A new scheme has been recently discussed, in which the
laser beam from the remote spacecraft illuminates a fiducial point
in the inertial sensor or spacecraft, rather than the proof mass.
The inertial sensor provides positioning and attitude reference,
and is also used to actuate the proof mass along the other degrees
of freedom. The proof mass could be monitored by employing
heterodyne laser metrology. This scheme, so-called separate
interferometry, has been adopted recently by LISA.\cite{Heinzel}
However LISA will not  fully exploit all its advantages as it may
still use two proof masses per spacecraft as references for two
different interferometer arms. ASTROD will employ one proof mass,
minimizing disturbances and simplifying control.

Replacing capacitive sensing with optical sensing, for drag-free
missions following LISA, has also been widely debated. Optical
sensing is more sensitive than electrostatic sensing and it
requires almost no coupling between the proof mass and
surroundings. Towards that direction many efforts to implement
optical sensing have been made in the last few
years,\cite{Acernese,KXSun,Speake,Xu,KXSunAng}
 but further laboratory
research is needed to develop a space qualified optical sensing
scheme. Light pressure could also be used for active control. A
more conservative design would combine an optical sensor and
capacitive active control.\cite{Clive} Larger gaps between the
proof mass and electrodes could then be used to minimize
disturbances.

Ultimately local gravitational gradients are  the limiting factor
for drag-free performance. In that context the influence of proof
mass geometries (spherical, cylindrical, cubic, polyhedral, etc)
on sensitivity and its repercussion for the overall inertial
sensor design merits further discussion. For monitoring and
correcting length changes due to thermal effects and slow
relaxations, ASTROD will use an absolute metrology
system.\cite{NiASR}

\subsection{Acceleration noise sources}

To discuss the acceleration noise, we consider a simplified
control loop model of a spacecraft and a single proof mass. The
acceleration noise is given by,\cite{Schumaker0,Schumaker,Sachie}

\begin{equation}\label{accnoise}
a_{n}\approx-K
X_{nr}+\frac{F_{str}}{m_{p}}+\left(\frac{F_{ns}+TN_{t}}{M}\right)\frac{K}{\omega^{2}u}
\end{equation}
where $F_{str}$ are stray forces directly acting on the proof mass
of mass $m_{p}$, $F_{ns}$ and $TN_{t}$ are external forces and
thruster force noise acting on the outer spacecraft of mass $M$.
$X_{nr}$ is the sensor readout sensitivity, $u$ is the control
loop gain and $\omega\equiv2\pi f$, where $f$ is the frequency.
External forces, thruster noise and readout sensitivity contribute
to acceleration noise because of the proof mass-spacecraft
coupling $K$. If the proof mass and spacecraft are highly
decoupled, then the acceleration noise will be given by stray
forces directly acting on the proof mass.

\begin{table}[t]
\tbl{Direct acceleration disturbances. The parameters are defined
as follows: $\lambda$ and $E$ denote cosmic ray impact rate and
incident energy, $P$ and $T$ are housing pressure and temperature,
respectively, $A_{P}$ the proof mass cross section, $\xi_{e}$ and
$\xi_{m}$ are the electrostatic and magnetic shielding factors,
respectively, $\delta T_{OB}$ describes optical bench temperature
fluctuations and $\xi_{TS}$ the thermal shielding factor between
optical bench and the proof mass, $\delta T_{SC}$ represents
spacecraft temperature fluctuations and $\alpha$ thermal expansion
coefficient. Finally $k_{B}$, $\sigma$ and $G$ denote the
Boltzman, Stefan-Boltzman and Newton Gravitational constants. The
rest of parameters are defined in the text.}
{\begin{tabular}{@{}cccc@{}} \hline
Environmental disturbances          &   \\
\hline
Cosmic rays                 & $f_{c}=\frac{\sqrt{2mE\lambda}}{m_{p}}$     \\
Residual gas      & $f_{rg}=\frac{\sqrt{2PA_{P}}}{m_{p}}(3k_{B}Tm_{N})^{1/4}$\\
Magnetic susceptibility I                 &$f_{m1}=\frac{2\chi}{\mu_{0}\rho\xi_{m}}\delta B_{SC}\nabla B_{SC}$  \\
Magnetic susceptibility II                & $f_{m2}=\frac{\sqrt{2}\chi}{\mu_{0}\rho\xi_{m}}\nabla B_{SC}\delta B_{ip}$  \\
Permanent magnetic moment   & $f_{m3}=\frac{1}{\sqrt{2}m_{p}\xi_{m}}M_{r}\nabla(\delta B)$  \\
Lorentz I    & $f_{L1}=\frac{v}{m_{p}\xi_{e}}q \delta B_{ip}$ \\
Lorentz II                          &$f_{L2}=\frac{v}{m_{p}\xi_{e}}B_{ip}\delta q$  \\
Radiometer effect
                          & $f_{re}=\frac{A_{p}P}{2m_{p}\xi_{TS}}\frac{\delta T_{OB}}{T}$  \\
Outgassing effect
                          & $f_{og}=10f_{re}$  \\
Thermal radiation pressure
                          & $f_{tp}=\frac{8\sigma}{3 m_{p}}\frac{A_{P}}{c}T^{3}\frac{\delta T_{OB}}{\xi_{TS}}$ \\
Gravity Gradients
                          & $f_{gg}=\frac{2GM}{r^{2}}\alpha\delta T_{SC}$
                          \\
                          \hline

\end{tabular} \label{ta1}}
\end{table}

In Table 1 direct acceleration noise sources with the exception of
sensor back action disturbances are listed.\cite{Schumaker}
Environmental disturbances can be divided into different groups
depending on their origin. There are disturbances caused by
impacts. Cosmic rays which penetrate the spacecraft shielding and
residual gas can deposit momentum onto the proof mass. There are
disturbances of magnetic origin. Proof mass magnetic
susceptibility, $\chi$, and residual permanent moment, $M_{r}$,
can interact with the residual local and/or the interplanetary
magnetic field, $B_{SC}$ and $B_{ip}$, respectively. There are
also disturbances associated with charge. Residual charge accrued
on the proof mass can interact with the interplanetary magnetic
field via Lorentz force. Finally, there are disturbances
associated with thermal fluctuations on the spacecraft. These
include the radiometer and outgassing effects, thermal radiation
pressure and gravitational gradients caused by thermally induced
spacecraft distortions.

Several key factors needed to quantify the total acceleration
noise, such as sensor readout noise, back action forces, and
stiffness terms, differ for different types of sensor. As an
example, optical sensing is nearly stiffness free, providing high
readout sensitivity with very low back action forces. On the other
hand, with the widely used capacitive sensor, high sensitivity is
achieved at the expense of increased acceleration noise and
stiffness. Also, because of the fact that close metallic surfaces
are needed (3-4 mm gaps in the case of LISA), other noise
contributions due to, for example, patch effects and dielectric
losses, become significant.

In the best scenario, the proof mass will ultimately be coupled to
the spacecraft by gravitational gradients. Force gradients of
electrostatic origin can be made negligible by implementing large
gaps between the proof mass and the surrounding metallic surfaces.
An analysis of proof mass geometries and thermal-gravitational
modeling of the spacecraft and payload are necessary to account
for gravitational disturbances and stiffness terms.

A preliminary quantitative analysis of acceleration noise
parameter requirements for  ASTROD is given in Refs.
\refcite{Pulido1} and \refcite{Pulido2}.

\subsection{Low-frequency acceleration noise sources}

The LISA observational bandwidth extends from 0.1 mHz to 0.1 Hz.
It has been pointed out that gravitational wave observations
extended to frequencies below 0.1 mHz, are desirable in the study
of certain astrophysical sources like massive black holes (MBH)
binaries at high redshift.\cite{Bender}

The ASTROD free falling proof masses will be separated by
distances of
 30 to 60 times longer than those of LISA. ASTROD
gravitational wave sensitivity curve will therefore be shifted to
lower frequencies than the target LISA bandwidth. At low
frequencies, spurious forces acting directly on the proof mass are
the dominant source of noise. This is the reason why, for a
mission like ASTROD, it is particularly important to identify
these sources of noise. An extended discussion of low-frequency
acceleration noise sources and low-frequency sensitivity curve for
gravitational waves for ASTROD is given in Ref. \refcite{Pulido1}.

Thermal, magnetic and electrostatic effects are sources of
low-frequency acceleration noise. Thermal noise arises due to
radiometer effect; fluctuating outgassing and thermal radiation
pressure assymetries; thermal distortion of the spacecraft and
residual gas impacts. The magnitude of these effects for a
particular mission are dependent on the mission orbit, which
dictates the thermal environment. Suppressing thermal disturbances
requires thermal diagnostics,\cite{Lobo} stable electronics,
passive and, in some cases, active thermal isolation, thermally
conductive electrodes (for the case of electrostatic
sensing/actuation) and high vacuum. A preliminary evaluation of
thermal disturbances for ASTROD shows that the outgassing effect,
thermal radiation pressure and thermally induced gravity gradients
could be at levels of about $f_{og}\approx$ 1.1 $\times$
10$^{-17}$ m s$^{-2}$ Hz$^{-1/2}$, $f_{tp}\approx$ 8 $\times$
10$^{-18}$ m s$^{-2}$ Hz$^{-1/2}$ and $f_{gg}\approx$ 5.4 $\times$
 10$^{-17}$ m s$^{-2}$ Hz$^{-1/2}$, respectively, at 0.1 mHz. A
vacuum pressure of the order of 10$^{-6}$ Pa, and a thermal
isolation factor, $\xi_{TS}$, of about 150, were assumed for these
estimates. Below 0.1 mHz, solar irradiance fluctuations become the
main cause of temperature fluctuations. Solar irradiance
fluctuations become more acute when approaching solar rotational
period, which is of the order of 25 days.\cite{Bender} This aspect
will be a crucial factor for the thermal diagnostic and thermal
isolation system design for ASTROD.

Magnetic noise at low frequencies is caused by interplanetary
magnetic field fluctuations, local gradients and magnetic field
fluctuations, eddy current damping, magnetic impurities, Lorentz
forces due to proof mass residual charge, etc. Suppression of
magnetic disturbances requires further reduction and shielding of
permanent magnets in the payload, improving magnetic shielding,
adopting a magnetic clean wiring (i.e, solar array rewiring, etc)
and power system. The most significant magnetic low-frequency
noise source is due to the interaction of proof mass magnetic
susceptibility with the interplanetary field fluctuations,
$f_{m2}$ (see Table 1). Assuming parameter values given in Ref.
\refcite{Pulido1}, $f_{m2}\approx2\times 10^{-17}$ m s$^{-2}$
Hz$^{-1/2}$ at 0.1 mHz. Disturbance $f_{m2}$ increases at low
frequencies as $f^{-2/3}$.

Other relevant noise sources at low frequencies are caused by
electrostatic effects, i.e., due to voltage noise, charge
fluctuations, DC voltages, dielectric losses, actuation noise,
etc. When employing capacitive sensing, the strategies to suppress
these noise sources are: increasing the separation between the
proof mass and the electrodes, active compensation of DC
voltages,\cite{Weber} avoidance of DC voltages applied to
drag-free degrees of freedom, high quality surface coatings to
minimize dielectric losses, high stability power supplies and
continuous discharging of the proof mass. Electrostatic noise is
caused by the capacitive sensor/actuator.

An obvious way for suppression of these noise sources is to
replace capacitive sensing by optical sensing. Nevertheless, other
concerns will arise if optical sensing is to be used. Thermal
distortion of optical components and changes in refractive index
with temperature modify optical paths. Because these noise sources
are due to thermal fluctuations, they will also be of significance
at low frequencies. New ideas addressing these problems, such as
using all reflective optics, by using gratings, has been
extensively discussed in the literature.\cite{KXSun} Diffractive
gratings have also been considered to enhance angular sensitivity,
compared with standard angular sensors based on laser
reflection.\cite{KXSunAng}

\subsection{Gravitational modelling}

Drag-free performance will ultimately be limited by local
gravitational fields and field gradients. Because of structural
distortion of the spacecraft due to thermal fluctuations, thermal
and gravitational disturbances need to be modelled together. For
the present discussion, we will be concerned only with the local
gravitational interaction between the test mass and a simplified
spacecraft structure.

We consider three different proof mass geometries: spherical,
cylindrical and cuboid. For simplicity, the spacecraft will be
assumed to be a hollow cylinder, as a first approximation.

Following appendix A, we can analyze the gravitational interaction
by means of inner, $q_{lm}$, and outer multipole moments,
$Q_{lm}$, that describe the proof mass, and the spacecraft and
payload mass distribution, respectively. Using expression
\ref{appB3}, the first non-zero outer multipole contribution of a
hollow cylinder is $Q_{20}$. Expression \ref{appB5} shows that
$Q_{20}$ couples to the proof mass inner multipoles $q_{00}$,
$q_{1\pm1}$, $q_{20}$, $q_{2\pm1}$ and $q_{2\pm2}$.

The first non-zero inner multipole moments for a parallelepiped
proof mass of sides 2a, 2b and 2c are $q_{00}=m_{p}/\sqrt{4\pi}$,
$q_{20}=2/3\sqrt{5/4\pi}m_{p}\left(2c^{2}-a^{2}-b^{2}\right)$ and
$q_{2\pm2}= 1/12\sqrt{15/2\pi}m_{p}(a^{2}-b^{2})$. On the other
hand for a cylindrical proof mass of radius $R$ and semi-height
$h$, we have $q_{00}= m_{p}/\sqrt{4\pi}$ and
$q_{20}=m_{p}\sqrt{5/4\pi}\left(h^{2}/3-R^{2}/4\right)$. Given
these values, it can be seen that the quadrupole moments vanish
for a cubic proof mass (as the one adopted by LISA). In the case
of ASTROD I, a proof mass of dimensions $50\times50\times35$ mm is
considered. In that case $q_{2\pm2}$ vanishes but not $q_{20}$. To
first order in the gravitational interaction with a cylindrical
tube the  $q_{20}$ term appears as a constant energy term and will
not contribute to the gravitational force. Disturbances
proportional to proof mass cross section area can then be
minimized by shortening one of the dimensions (as is the case of
ASTROD I proof mass). A trade off between the acceleration
disturbances which are proportional to proof mass cross sectional
area, and the gravitational interaction needs to be done. If a
cylindrical proof mass is utilized we can also suppress quadrupole
gravitational interaction by choosing $h/R=\sqrt{3}/2$. In the
same way disturbances proportional to cross sectional area can be
suppressed by shortening $h$, without altering gravitational
interaction with "far away" gravitational asymmetries. A
preliminary analysis of gravitational force gradients for ASTROD I
is given in Ref. \refcite{GRAVSachie}.

Another issue that needs to be considered for ASTROD is the fact
that the relative distances and angles between the spacecraft are
not constant along their orbits. Therefore, telescopes utilized
for laser beam pointing need to be steered during the mission. For
active gravitational compensation ASTROD will employ dummy
telescopes.\cite{NiGRG,Pulido2}

\subsection{Back action disturbances}

When deciding which type of sensor to use, there are two main
options to consider. First, we could consider a low-stiffness
sensor. In that case, the proof mass is highly decoupled from the
spacecraft, at the expense of losing sensitivity. The other option
is to employ a high-stiffness sensor to achieve better readout
sensitivity, at the expense of a high level of coupling between
the proof mass and the surrounding structure. Capacitive sensing
exemplifies this issue. To improve sensitivity we need to place
the electrodes closer to the proof mass. By doing that, the
stiffness and back action disturbances increase (see Table 2).

To understand the principle behind stiffness and back action
disturbances due to electrostatic sensing and actuation, we first
consider the total mechanical energy for a capacitive sensor-proof
mass system. Following Refs. \refcite{Schumaker0} and
\refcite{Sachie}, the total mechanical energy is given by

\begin{equation}
W=-\frac{1}{2}\sum_{i}C_{i}(V_{i}-V_{s})^{2}+
\frac{1}{2}\frac{q^{2}}{C}+qV_{s}
\end{equation}\label{electrostaticenergy}
where $q$ is the net charge of the proof mass; $C$ is the
coefficient of capacitance of the proof mass: $C=\sum_{i}C_{i}$,
where $i$ = $x_{1}, x_{2}, y_{1}, y_{2}, z_{1}, z_{2}, g$ defines
the capacitances formed by an electrode facing a proof mass side,
and capacitance to ground, $C_{g}$; $V_{s}$ is the voltage induced
on the proof mass due to the applied voltages, $V_{i}$, and
voltage to ground, $V_{g}$: $V_{s}=C^{-1}\sum_{i}C_{i}V_{i}$.

The force acting along a generic direction, assuming neither
charge and voltage gradients, is given by

\begin{equation}
F=\frac{1}{2}\sum_{i}C'_{i}(V_{i}-V_{M})^{2}
\end{equation}
where $V_{M}=V_{s}+\frac{q}{C}$, and $C'_{i}$ is the derivative of
capacitance $C_{i}$ along the generic direction.

Force disturbances can then be considered of two types: a)
position dependent disturbances, caused by fluctuating position
and attitude of the proof mass (stiffness terms), and b) position
independent disturbances, caused by voltage and charge
fluctuations. Along a generic sensitive (drag-free) axis, we can
write the fluctuating force terms due to charge and voltage
fluctuations as

\begin{equation}
\delta F_{\delta V,1}=\left(\sum_{i}
C_{i}'(V_{i}-V_{s})-V_{m_{1}}'C\right)\delta V_{i}
\end{equation}

\begin{equation}
\delta F_{\delta
V,2}=\left(\sum_{i}C_{i}'\frac{q}{C}+V_{m_{2}}'\sum_{i}
C_{i}\right)\delta V_{i}
\end{equation}

\begin{equation}
\delta F_{\delta q,1}=V_{m_{1}}'\delta q
\end{equation}

\begin{equation}
\delta F_{\delta q,2}=V_{m_{2}}'\delta q
\end{equation}
where

\begin{equation}
V_{m_{1}}'=\frac{1}{C}\sum_{i}C_{i}'(V_{i}-V_{s})
\end{equation}
and
\begin{equation}
V_{m_{2}}'=-\frac{q}{C^{2}}\sum_{i}C_{i}'
\end{equation}

Table 2 shows the back action disturbances given above for the
special case of a one dimensional capacitive sensor and one
translational degree of freedom.\cite{Schumaker} Disturbances due
to dielectric losses and patch fields are also listed. An extended
discussion of electrostatic back action disturbances for ASTROD I
and ASTROD can be found in Ref. \refcite{Sachie} and
\refcite{Pulido1}, respectively.

Position dependent disturbances (stiffness terms)   can be
obtained by calculating the variations in capacitances, $\delta
C_{i}$, and capacitance gradients, $\delta C_{i}'$.

In appendix B formulae for capacitances, capacitance gradients and
their fluctuations, given by $C_{i}$, $C_{i}'$, $\delta C_{i}$ and
$\delta C_{i}'$, respectively, are obtained for the special case
of a 6-degree of freedom, cubic, capacitive sensor. These
expressions codify cross-coupling effects between translational
and rotational degrees of freedom.

On the other hand, optical sensing offers advantages in terms of
high sensitivity and low back action forces.  Optical readout
sensitivity is ultimately limited by laser shot noise. Laser shot
noise is proportional to $P^{-1/2}$, where $P$ is the laser power.
Lasing back action force is proportional to the laser power, and
is given by $2P/c$, where $c$ is the speed of light. This force
can, in principle, be compensated to a high degree of accuracy.

\begin{table}[t]
\tbl{Sensor back action acceleration disturbances. The parameters
are defined as follows: $\delta v_{diel}$ denotes voltage
fluctuation due to dielectric losses, $V_{0}$ is dc bias voltage,
$V_{pe}$ average patch potential, $V_{x0}$ average potential
across opposite side of sensor, $\delta V_{d}$ fluctuations in
voltage different across opposite side of sensor and $d$ is the
gap between the proof mass and surrounding housing and electrodes.
The rest of parameters are defined in the text.}
{\begin{tabular}{@{}cccc@{}} \hline
Back action disturbances \\
\hline
Dielectric losses                 & $f_{DL}=\frac{\sqrt{2}C_{x}}{m_{p}d}V_{0}\delta v_{diel}$   \\
Patch fields (Uncompensated) & $f_{pe}=\frac{1}{m_{p}d}\frac{C_{x}}{C}V_{pe}\delta q$\\
$\delta V_{d}\times V_{0g}$      & $f_{\delta V,1}=\frac{C_{x}}{m_{p}d}\frac{C_{x}}{C}(V_{x0}-V_{g})\delta V_{d}$  \\
$\delta V_{d}\times q$       & $f_{\delta V,2}=\frac{q}{m_{p}d}\frac{C_{x}}{C}\delta V_{d}$   \\
$\delta q \times V_{d}$       & $f_{\delta q,1}=\frac{1}{m_{p}d}\frac{C_{x}}{C} V_{d}\delta q$  \\
$\delta q\times q$       & $f_{\delta q,2}=\frac{q}{m_{p}d^{2}}\frac{C_{x}}{C^{2}}\Delta d\delta q$  \\
\hline
\end{tabular} \label{ta1}}
\end{table}

\section{Charging disturbances}

Galactic cosmic rays (GCR) and solar energetic particles (SEP)
incident on the spacecraft will result in the accumulation of
charge on the proof mass. Charge accrued on the proof mass leads
to numerous sources of noise. Some of these noise sources have
been described and discussed above. First, when summarizing direct
acceleration noise sources, the Lorentz force due to the movement
of the charged proof mass through the interplanetary magnetic
field was discussed. We have also discussed, in the particular
case of employing capacitive sensing/actuation, how the proof mass
charge couples to sensing/actuation voltages to induce spurious
forces and stiffness terms that will affect the performance of the
inertial sensor.

Disturbances associated with charging can be divided into three
types: a) those which are proportional to charge accrued, $q$, b)
those proportional to $\delta q$, which are so-called "shot noise"
terms and c) mixed terms proportional to $q\delta q$. This
division is of importance in understanding disturbance suppression
schemes (see discussion below).

Time dependent forces also contribute to the spectral noise
density. Proof mass charging is a time dependent process and both
Coulomb and Lorentz forces give rise to coherent Fourier signals
(CHS). Assuming a linear increase of proof mass charge with time,
the total charge can be written as $q(t)= \bar{\dot{q}}t+\delta
q$, where $\bar{\dot{q}}$ is the mean charging rate. Following
Ref. \refcite{Shaul}  the acceleration noise terms due to Coulomb
and Lorentz interactions are given by

\begin{equation}
a_{CHS}= h_{k}(t)=\left(\phi_{k}+\Theta_{k}\right)t+\Xi_{k}t^{2}
\end{equation}
where

\begin{equation}\label{coherentonedim}
\phi_{k}=\frac{\bar{\dot{q}}vB_{ip}}{m_{p}\xi_{e}},
\hspace{0.2cm}\Theta_{k}=\frac{\bar{\dot{q}}C_{x}}{m_{p}Cd}V_{d},\hspace{0.1cm}\textrm{and}
\hspace{0.2cm}\Xi_{k}=\frac{2C_{x}}{m_{p}}\left(\frac{\bar{\dot{q}}}{Cd}\right)^{2}
\Delta d
\end{equation}
using a parallel plate approximation to estimate capacitances and
capacitance derivatives. The parameters used above are defined as
follows: $v$ is the orbital velocity of the proof mass, $B_{ip}$
is the interplanetary magnetic field, $\xi_{e}$ is the
electrostatic shielding factor, $C_{x}$ is the capacitance along
the sensitive axis, $V_{d}$ is the voltage difference between
opposite sensor sides, and $d$ and $\Delta d$ are the capacitance
gap and gap asymmetry, respectively.

Inspection of (\ref{coherentonedim}) shows that geometrical and
electrostatic asymmetries in the inertial sensor contribute to the
appearance of these signals. Geometrical asymmetry arises due to
limited machining accuracy and it is represented by an asymmetry
in the capacitance gap. Electrostatic asymmetry is due to a stray
DC potential imbalance between opposite sides of the sensor. These
residual DC stray potentials are dependent on the work function of
the metallic surfaces. These potentials are measurable in average,
for each electrode, and can be balanced by appropriate applied
bias voltages.\cite{Weber} Ultimately, voltage offset compensation
will depend on voltage measurement precision required and work
function domains stability in periods of time  comparable with the
measurement integration time. In the context of LISA mission,
these coherent Fourier signals can exceed the instrumental noise
target, for typical parameter values.\cite{Shaul}

At low frequencies, charging disturbances and coherent charging
signals are of particular concern. Those so-called "shot noise"
charging disturbances and coherent signals scale, roughly
speaking, with frequency as $1/f$.\cite{Shaul} Of special note is
the acceleration disturbance proportional to residual voltage
difference across opposite sides of the sensor and charge
fluctuations, $f_{\delta q,1}$ (see Table 2).  By active
compensation, the potential difference across opposite sensor
sides can be balanced to $\sim$1 mV.\cite{Weber} To improve on the
LISA acceleration noise by a factor of 10 at 0.1 mHz, ASTROD would
need to reduce this value to below 0.7 mV,\cite{Pulido2} giving a
disturbance level $f_{\delta q,1}\approx$ 1.4 $\times$ $10^{-16}$
m s$^{-2}$ Hz$^{-1/2}$.

\subsection{Discharging schemes}

When discussing discharging schemes we have to keep in mind that
not only charge accrued by the proof mass but also the charging
rate are potential causes of noise.\cite{Shaul2,KXSun2}
Discharging the proof mass will suppress some charging
disturbances and will reduce the coupling of the proof mass with
its surroundings. Coherent signals are proportional to mean
charging rate. By "mean charging rate" we mean that charging and
discharging rates are added linearly to give a net rate. Therefore
by accurately matching charging and discharging rates, these
signals can be suppressed.\cite{Shaul} Nevertheless the fact that
proof mass discharging involves the transport of charge
"packages", will cause additional shot noise. Shot noise due to
proof mass charging and discharging cannot partially cancel each
other because the charging and discharging processes are
statistically independent. Therefore their respective shot noise
terms have to be added quadratically.

For LISA, a continuous discharging scheme will be adopted. The
continuous discharging process will consist basically of two
steps. Firstly, the proof mass charge needs to be accurately
measured. That can be done by applying a sinusoidal dither voltage
and measuring the displacement along a non drag-free axis. This
proof mass displacement is proportional to proof mass charge.
Secondly, UV light will shine on the proof mass and/or surrounding
electrodes to discharge the proof mass via the photoelectric
effect.\cite{Shaul,Shaul2,KXSun2}

The ASTROD strategy to suppress charging noise will depend on
which sensing/actuation device is employed. ASTROD could benefit
from the replacement of capacitive sensing by optical sensing.
Even if a capacitive scheme is employed for force actuation,
charging disturbances and coherent signals could be suppressed by
increasing the gaps between the test mass and surrounding
surfaces. If optical force actuation is employed, then only
Lorentz type disturbances need to be considered. In that case, the
charging requirements can be relaxed and the discharging scheme
can be simplified.

\section{Inertial sensors for other missions to test fundamental physics}

Follow on LISA mission concepts have been proposed, not only to
extend the observational bandwidth to lower frequencies, but also
to fill the gap between space antennae and ground based
interferometric facilities (LIGO, GEO600, VIRGO, Advanced LIGO,
LCGT, etc). Follow-on mission concepts include the Big Bang
Observer (BBO) and the Japanese antenna DECi-hertz Interferometer
Gravitational wave Observer (DECIGO). Both BBO and DECIGO are to
be designed to have an optimum sensitivity between 0.1 Hz and 10
Hz. The objectives of the decihertz antennae are: measuring the
expansion rate of the universe; determining the equation of state
of dark energy, by observing the coalescence of binary neutron
stars and stellar mass black holes; and shedding light on the
growth of supermassive black holes, by studying the merger of
intermediate mass black holes.\cite{BBO,DECIGO}

These missions will also measure the relative distance, or
maintain the distance by feedback, between test masses in nearly
free fall. To achieve their objectives they need to reduce
residual spurious forces to approximately a hundredth of the LISA
goal.

These missions are conceptually different. To shift the
gravitational wave sensitivity curve towards the decihertz level,
the effective interferometer arm length has to be, approximately,
a hundredth of the LISA arm length (5 $\times$ 10$^{6}$ km). The
BBO preliminary conceptual design consists of a constellation of
spacecraft with a LISA-type design and an arm length of 5 $\times$
10$^{4}$ km. On the other hand, DECIGO plans to place in space a
Fabry-Perot cavity of length 1000 km and finesse of approximately
10. To accomplish this, DECIGO would place massive mirrors of 100
kg mass and 1 meter diameter in free fall. Large actuation forces,
to keep the cavity in resonance, will be required. This condition
and the stringent residual acceleration noise levels, of the order
of 4 $\times$ 10$^{-19}$ m s$^{-2}$ Hz$^{-1/2}$, seem to be
difficult to reconcile. Nevertheless one could employ a large
control loop gain to minimize the acceleration disturbances due to
actuation forces. Other technological difficulties common to both
missions are related to the requirement for low residual gas
pressure. Extremely low pressure could be achieved by venting some
of the gas to outer space. However, this could cause other
problems such as drag of the proof mass because of residual gas
flow, and undesired particles coming from the thruster propellant,
brought into the proof mass housing.

\section*{Acknowledgments}

The author thanks W.-T. Ni for his useful comments on this work
and the manuscript and D. N. Shaul for discussing issues related
to charge management. This work was funded by the National Natural
Science Foundation (Grant No 10475114) and the Foundation of Minor
Planets.

\appendix{\section{ Gravitational interaction in terms of multipole
moments}}

The gravitational potential can be written in terms of multipole
moments as\cite{GRAVSachie,Jackson}

\begin{equation}
V = -4\pi G
\sum_{l=0}^{\infty}\sum_{m=-l}^{l}\frac{1}{2l+1}q_{lm}Q_{lm}
\label{appB1}
\end{equation}
where $q_{lm}$ ($Q_{lm}$) are the inner (outer) moments defined
respectively by

\begin{equation}
q_{lm}=\int_{v_{t}}\rho_{t}(\vec{x}')r'^{l}Y^{*}_{lm}(\theta',\phi')d^{3}\vec{x}'\label{appB2}
\end{equation}
and

\begin{equation}
Q_{lm}=\int_{v_{s}}\rho_{s}(\vec{x})r^{-(l+l)}Y_{lm}(\theta,\phi)d^{3}\vec{x}
\label{appB3}
\end{equation}

If the inner multipoles of the test mass, $q_{lm}$, are known in a
given reference frame, then the inner multipoles with respect to a
new reference frame, can be obtained. Let us consider a position
$\vec{r}"=\vec{r}'+\vec{r}$. The test mass position with respect
to the reference frame, of origin O, in which the multipoles are
known are denoted by $\vec{r}$. On the other hand, $\vec{r}'$
denotes the position vector of the origin O, with respect to the
new reference frame in which we wish to work out the multipoles.
In the special case of pure translations,\cite{D'Urso}

\begin{eqnarray}
r"^{L}Y^{*}_{LM}(\theta,\phi)=
\sum_{l,l'=0}^{L}\sum_{m,m'}\sqrt{\frac{4\pi(2L+1)!}{(2l'+1)!(2l+1)!}}r'^{l'}r^{l}\times\nonumber\\
\delta_{L,l+l'}C(l',m',l,m,L,M)Y^{*}_{l'm'}(\theta',\phi')
Y_{lm}(\theta,\phi) \label{appB4}
\end{eqnarray}

Using eq. \ref{appB4} we can rewrite the gravitational potential
energy \ref{appB1} in terms of the known inner multipoles and the
translational parameters between the two reference frames as

\begin{eqnarray}
V(\vec{r})= -4\pi
G\sum_{L=0}^{\infty}\sum_{M=-L}^{L}\frac{1}{2L+1}Q_{LM}\times\nonumber\\
\sum_{l,l'=0}^{L}\sum_{m,m'}\sqrt{\frac{4\pi(2L+1)!}{(2l'+1)!(2l+1)!}}C(l',m',l,m,L,M)\delta_{L,l+l'}r^{l'}Y^{*}_{l'm'}(\theta,\phi)
q_{lm} \label{appB5}
\end{eqnarray}

The gravitational force can be then easily calculated by taking
the derivatives of \ref{appB5} with respect to the translational
parameters.

\section{A cubical inertial sensor. Capacitance calculations}

We consider a six-dimensional degree of freedom model for the
capacitive sensing/actuation device. A cuboid proof mass of side
lengths $(2L_{x}, 2L_{y}, 2L_{z})$ is inserted into a three
dimensional capacitive sensor. The gaps at the equilibrium
position between the proof mass and the electrodes are denoted by
$(D_{x}, D_{y}, D_{z})$. The proof mass translational degrees of
freedom are denoted by $(d_{x}, d_{y}, d_{z})$. The proof mass
rotational degrees of freedom are given by the Euler angles
$(\phi, \theta, \psi)$, in the so-called "x-convention".

We first consider the electrostatic energy density to work out
capacitances for this model. If the electric field between two
conducting surfaces is defined by $\vec{E}$, then the
electrostatic energy density is given by
$\omega=\frac{1}{2}\varepsilon_{0}E^{2}$. The proof mass faces and
the electrodes will define the capacitances. In the case in which
the displacement and rotation of the proof mass are infinitesimal,
we can approximate the electric field between conducting surfaces
by $\vec{E}_{i}\simeq\frac{(V_{M}-V_{i})}{\triangle
x_{i}}\vec{u}_{i}$, where $V_{M}$ ($V_{i}$) and $\triangle x_{i}$
denote the proof mass (electrode) potential and the capacitance
gap in the $i$-direction, respectively. The electrostatic energy
is then given by

\begin{equation}
W^{\pm}(d_{x}, d_{y}, d_{z}, \phi, \theta,
\psi)\simeq\frac{1}{2}\varepsilon_{0}(V_{M}-V_{i})^{2}\int\frac{dV}{\triangle
x_{i}^{\pm 2}}
\end{equation}
where by the simbol $\pm$ we differentiate between the gaps at
opposite sides of the sensor.

By integrating along the gap (in this case we choose the gap along
the z-axis), the electrostatic energy can be written as

\begin{equation}
W^{\pm}(d_{x}, d_{y}, d_{z}, \phi, \theta,
\psi)\simeq-\frac{1}{2}\varepsilon_{0}(V_{M}-V_{i})^{2}\int\frac{dxdy}{\triangle
x_{i}^{\pm}(x,y,d_{x}, d_{y}, d_{z}, \phi, \theta, \psi)}
\end{equation}
and we can define the capacitances by

\begin{equation}
C_{x_{i}}^{\pm}(d_{x}, d_{y}, d_{z}, \phi, \theta,
\psi)\simeq\varepsilon_{0}\int dx_{j}dx_{k}\frac{1}{\triangle
x_{i}^{\pm}(x_{j},x_{k},d_{x}, d_{y}, d_{z}, \phi, \theta,
\psi)}\label{cap1}
\end{equation}

The information about how different degrees of freedom couple to
each other is codified in \ref{cap1}.

To explicitly work out capacitances, we translate and rotate the
proof mass by the parameters $(d_{x}, d_{y}, d_{z}, \phi, \theta,
\psi)$.

The rotation matrix in terms of the Euler angles $(\phi, \theta,
\psi)$ is written as
\begin{eqnarray}
N(\phi, \theta, \psi)=\hspace{5cm}\nonumber\\
\left(%
\begin{array}{ccc}
  \cos \psi \cos \phi -\cos \theta \sin \phi \sin \psi & \cos \psi \sin \phi+\cos \theta \cos \phi \sin \psi & \sin \psi \sin\theta \\
  -\sin \psi \cos \phi-\cos \theta \sin \phi \cos \psi & -\sin \psi \sin \phi+\cos \theta \cos \phi \cos \psi & \cos \psi \sin \theta \\
  \sin \theta \sin \phi & -\sin \theta \cos \phi & \cos \theta \\
\end{array}%
\right)
\end{eqnarray}

Given this matrix we can define the normal vectors to the proof
mass faces after a $(\phi, \theta, \psi)$-rotation, with respect
to a fixed reference frame. This reference frame has its origin in
the geometrical center of the sensor and its axis orthogonal to
electrode surfaces. The normal vectors can be written as

\begin{equation}
n_{1}=\left(%
\begin{array}{c}
  n_{1x} \\
  n_{1y} \\
  n_{1z} \\
\end{array}%
\right)=\left(%
\begin{array}{c}
   \cos \psi \cos \phi -\cos\theta \sin \phi \sin \psi\\
   \cos \psi \sin \phi+\cos \theta \cos \phi \sin \psi\\
   \sin \psi \sin \theta\\
\end{array}%
\right)
\end{equation}

\begin{equation}
n_{2}=\left(%
\begin{array}{c}
  n_{2x} \\
  n_{2y} \\
  n_{2z} \\
\end{array}%
\right)=\left(%
\begin{array}{c}
   -\sin \psi \cos \phi-\cos \theta \sin \phi \cos \psi \\
   -\sin \psi \sin \phi+\cos \theta \cos \phi \cos \psi\\
   \cos \psi \sin \theta\\
\end{array}%
\right)
\end{equation}

\begin{equation}
n_{3}=\left(%
\begin{array}{c}
  n_{3x} \\
  n_{3y} \\
  n_{3z} \\
\end{array}%
\right)=\left(%
\begin{array}{c}
  \sin \theta \sin \phi \\
  -\sin \theta \cos \phi\\
  \cos \theta\\
\end{array}%
\right)
\end{equation}

The equations of the proof mass faces, are defined by
\begin{equation}
\left(\vec{x}-\vec{P}_{i}\right)\cdot \vec{n}_{i}=0
\end{equation}
where $\vec{P}_{i}=\pm L_{i}n_{i}+\vec{d}$, being
$\vec{d}\equiv(d_{x}, d_{y}, d_{z})$ a displacement, and $L_{i}$
defines the semi-length of the proof mass along the three axes x,
y and z.

Using this expressions we can obtain the capacitance gaps for the
three directions. These are given by

\begin{equation}
\triangle
x^{\pm}=F^{\pm}_{x}\pm\left(y+A_{x}^{\pm}\right)n_{1}^{x}\pm\left(z+B_{x}^{\pm}\right)n_{2}^{x}
\end{equation}

\begin{equation}
\triangle
y^{\pm}=F^{\pm}_{y}\pm\left(x+A^{\pm}_{y}\right)n_{1}^{y}\pm\left(z+B^{\pm}_{y}\right)n_{2}^{y}
\end{equation}

\begin{equation}
\triangle
z^{\pm}=F_{z}^{\pm}\pm\left(x+A_{z}^{\pm}\right)n_{1}^{z}\pm\left(y+B_{z}^{\pm}\right)n_{2}^{z}
\end{equation}

\begin{figure}[t]
\centering
% Use the relevant command for your figure-insertion program
% to insert the figure file.
% For example, with the option graphics use
\includegraphics[height=8cm]{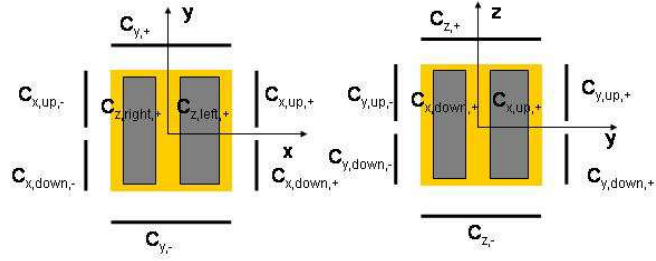}
%
% If not, use
%\picplace{5cm}{2cm} % Give the correct figure height and width in cm
%
\vspace*{-55pt}
 \caption{Schematic view of capacitive sensing electrodes.}
\label{CS}       % Give a unique label
\end{figure}

We define the capacitances as

\begin{equation}
C_{x,up}^{\pm}=\varepsilon_{0}\int_{0}^{L_{y}}
dy\int_{-L_{z}}^{L_{z}}dz\frac{1}{\triangle x^{\pm}(y,z)}
\end{equation}

\begin{equation}
C_{x,down}^{\pm}=\varepsilon_{0}\int_{-L_{y}}^{0}
dy\int_{-L_{z}}^{L_{z}}dz\frac{1}{\triangle x^{\pm}(y,z)}
\end{equation}

\begin{equation}
C_{y,up}^{\pm}=\varepsilon_{0}\int_{-L_{x}}^{L_{x}}
dx\int_{0}^{L_{z}}dz\frac{1}{\triangle y^{\pm}(y,z)}
\end{equation}

\begin{equation}
C_{y,down}^{\pm}=\varepsilon_{0}\int_{-L_{x}}^{L_{x}}
dx\int_{-L_{z}}^{0}dz\frac{1}{\triangle y^{\pm}(y,z)}
\end{equation}

\begin{equation}
C_{z,left}^{\pm}=\varepsilon_{0}\int_{0}^{L_{x}}
dx\int_{-L_{y}}^{L_{y}}dy\frac{1}{\triangle z^{\pm}(x,y)}
\end{equation}

\begin{equation}
C_{z,right}^{\pm}=\varepsilon_{0}\int_{-L_{x}}^{0}
dx\int_{-L_{y}}^{L_{y}}dy\frac{1}{\triangle z^{\pm}(x,y)}
\end{equation}
where by up, down, left, right, we indicate that two electrodes
face each proof mass face (see Fig. \ref{CS}).

The capacitances in the x-direction are given by

\begin{eqnarray}
C_{x,up}^{+}=\frac{\varepsilon_{0}}{n_{2}}L_{y}\ln\frac{F_{x}^{+}+(L_{z}+B_{x}^{+})n^{x}_{2}+(A_{x}^{+}+L_{y})n^{x}_{1}}{F_{x}^{+}+(B_{x}^{+}-L_{z})n^{x}_{2}+(A_{x}^{+}+L_{y})n^{x}_{1}}\nonumber\\
-\frac{\varepsilon_{0}}{n^{x}_{2}}\frac{F_{x}^{+}+(L_{z}+B_{x}^{+})n^{x}_{2}+A_{x}^{+}n^{x}_{1}}{n^{x}_{1}}\ln\frac{F_{x}^{+}+(L_{z}+B_{x}^{+})n^{x}_{2}+(A_{x}^{+}+L_{y})n^{x}_{1}}{F_{x}^{+}+(L_{z}+B_{x}^{+})n^{x}_{2}+A_{x}^{+}n^{x}_{1}}\nonumber\\
+\frac{\varepsilon_{0}}{n^{x}_{2}}\frac{F_{x}^{+}+(B_{x}^{+}-L_{z})n^{x}_{2}+A_{x}^{+}n^{x}_{1}}{n^{x}_{1}}\ln\frac{F_{x}^{+}+(B_{x}^{+}-L_{z})n^{x}_{2}+(A_{x}^{+}+L_{y})n^{x}_{1}}{F_{x}^{+}+(B_{x}^{+}-L_{z})n^{x}_{2}+A_{x}^{+}n^{x}_{1}}
\end{eqnarray}

\begin{eqnarray}
C_{x,up}^{-}=-\frac{\varepsilon_{0}}{n^{x}_{2}}L_{y}\ln\frac{F_{x}^{-}-(L_{z}+B_{x}^{-})n^{x}_{2}-(A_{x}^{-}+L_{y})n^{x}_{1}}{F_{x}^{-}-(B_{x}^{-}-L_{z})n^{x}_{2}-(A_{x}^{-}+L_{y})n^{x}_{1}}\nonumber\\
-\frac{\varepsilon_{0}}{n^{x}_{2}}\frac{F_{x}^{-}-(L_{z}+B_{x}^{-})n^{x}_{2}-A_{x}^{-}n^{x}_{1}}{n^{x}_{1}}\ln\frac{F_{x}^{-}-(L_{z}+B_{x}^{-})n^{x}_{2}-(A_{x}^{-}+L_{y})n^{x}_{1}}{F_{x}^{-}-(L_{z}+B_{x}^{-})n^{x}_{2}-A_{x}^{-}n^{x}_{1}}\nonumber\\
+\frac{\varepsilon_{0}}{n^{x}_{2}}\frac{F_{x}^{-}-(B_{x}^{-}-L_{z})n^{x}_{2}-A_{x}^{-}n^{x}_{1}}{n^{x}_{1}}\ln\frac{F_{x}^{-}-(B_{x}^{-}-L_{z})n^{x}_{2}-(A_{x}^{-}+L_{y})n^{x}_{1}}{F_{x}^{-}-(B_{x}^{-}-L_{z})n^{x}_{2}-A_{x}^{-}n^{x}_{1}}
\end{eqnarray}

\begin{eqnarray}
C_{x,down}^{+}=\frac{\varepsilon_{0}}{n^{x}_{2}}L_{y}\ln\frac{F_{x}^{+}+(L_{z}+B_{x}^{+})n^{x}_{2}+(A_{x}^{+}-L_{y})n^{x}_{1}}{F_{x}^{+}+(B_{x}^{+}-L_{z})n^{x}_{2}+(A_{x}^{+}-L_{y})n^{x}_{1}}\nonumber\\
+\frac{\varepsilon_{0}}{n^{x}_{2}}\frac{F_{x}^{+}+(L_{z}+B_{x}^{+})n^{x}_{2}+A_{x}^{+}n^{x}_{1}}{n^{x}_{1}}\ln\frac{F_{x}^{+}+(L_{z}+B_{x}^{+})n^{x}_{2}+(A_{x}^{+}-L_{y})n^{x}_{1}}{F_{x}^{+}+(L_{z}+B_{x}^{+})n^{x}_{2}+A_{x}^{+}n^{x}_{1}}\nonumber\\
-\frac{\varepsilon_{0}}{n^{x}_{2}}\frac{F_{x}^{+}+(B_{x}^{+}-L_{z})n^{x}_{2}+A_{x}^{+}n^{x}_{1}}{n^{x}_{1}}\ln\frac{F_{x}^{+}+(B_{x}^{+}-L_{z})n^{x}_{2}+(A_{x}^{+}-L_{y})n^{x}_{1}}{F_{x}^{+}+(B_{x}^{+}-L_{z})n^{x}_{2}+A_{x}^{+}n^{x}_{1}}
\end{eqnarray}

\begin{eqnarray}
C_{x,down}^{-}=-\frac{\varepsilon_{0}}{n^{x}_{2}}L_{y}\ln\frac{F_{x}^{-}-(L_{z}+B_{x}^{-})n^{x}_{2}-(A_{x}^{-}-L_{y})n^{x}_{1}}{F_{x}^{-}-(B_{x}^{-}-L_{z})n^{x}_{2}-(A_{x}^{-}-L_{y})n^{x}_{1}}\nonumber\\
+\frac{\varepsilon_{0}}{n^{x}_{2}}\frac{F_{x}^{-}-(L_{z}+B_{x}^{-})n^{x}_{2}-A_{x}^{-}n^{x}_{1}}{n^{x}_{1}}\ln\frac{F_{x}^{-}-(L_{z}+B_{x}^{-})n^{x}_{2}-(A_{x}^{-}-L_{y})n^{x}_{1}}{F_{x}^{-}-(L_{z}+B_{x}^{-})n^{x}_{2}-A_{x}^{-}n^{x}_{1}}\nonumber\\
-\frac{\varepsilon_{0}}{n^{x}_{2}}\frac{F_{x}^{-}-(B_{x}^{-}-L_{z})n^{x}_{2}-A_{x}^{-}n^{x}_{1}}{n^{x}_{1}}\ln\frac{F_{x}^{-}-(B_{x}^{-}-L_{z})n^{x}_{2}-(A_{x}^{-}-L_{y})n^{x}_{1}}{F_{x}^{-}-(B_{x}^{-}-L_{z})n^{x}_{2}-A_{x}^{-}n^{x}_{1}}
\end{eqnarray}
where

\begin{eqnarray}
F_{x}^{\pm}\equiv L_{x}(1-n_{1x})+D_{x}\mp d_{x}\\
A_{x}^{\pm}\equiv \mp L_{x}n_{1y}-d_{y}\\
B_{x}^{\pm}\equiv\mp L_{x}n_{1z}-d_{z}\\
n^{x}_{1}\equiv\frac{n_{1y}}{n_{1x}}\\
n^{x}_{2}\equiv\frac{n_{1z}}{n_{1x}}
\end{eqnarray}

The capacitances in the y-direction are

\begin{eqnarray}
C_{y,up}^{\pm}=C_{x,up}^{\pm}((L_{y},L_{z},A_{x}^{\pm},B_{x}^{\pm},n^{x}_{1},n^{x}_{2})\rightarrow
(L_{z},L_{x},B_{y}^{\pm},A_{y}^{\pm},n^{y}_{2},n^{y}_{1}))\\
C_{y,down}^{\pm}=C_{x,down}^{\pm}((L_{y},L_{z},A_{x}^{\pm},B_{x}^{\pm},n^{x}_{1},n^{x}_{2})\rightarrow
(L_{z},L_{x},B_{y}^{\pm},A_{y}^{\pm},n^{y}_{2},n^{y}_{1}))
\end{eqnarray}
where

\begin{eqnarray}
F_{y}^{\pm}\equiv L_{y}(1-n_{2y})+D_{y}\mp d_{y}\\
A_{y}^{\pm}\equiv \mp L_{y}n_{2x}-d_{x}\\
B_{y}^{\pm}\equiv\mp L_{y}n_{2z}-d_{z}\\
n^{y}_{1}\equiv\frac{n_{2x}}{n_{2y}}\\
n^{y}_{2}\equiv\frac{n_{2z}}{n_{2y}}
\end{eqnarray}

The capacitances in the z-direction are given by

\begin{eqnarray}
C_{z,left}^{\pm}=C_{x,up}^{\pm}(L_{z}\rightarrow L_{y};L_{y}\rightarrow L_{x})\\
C_{z,right}^{\pm}=C_{x,down}^{\pm}(L_{z}\rightarrow
L_{y};L_{y}\rightarrow L_{x})
\end{eqnarray}
where now

\begin{eqnarray}
F_{z}^{\pm}\equiv L_{z}(1-n_{3z})+D_{z}\mp d_{z}\\
A_{z}^{\pm}\equiv \mp L_{z}n_{3x}-d_{x}\\
B_{z}^{\pm}\equiv\mp L_{z}n_{3y}-d_{y}\\
n^{z}_{1}\equiv\frac{n_{3x}}{n_{3z}}\\
n^{z}_{2}\equiv\frac{n_{3y}}{n_{3z}}
\end{eqnarray}

We can approximate the expressions of capacitances by taking into
account that $F\gg$ $A$ and $B$. Then we have

\begin{equation}
C_{x,up}^{+}=2\varepsilon_{0}L_{y}L_{z}\frac{1}{F_{x}^{+}+B_{x}^{+}n^{x}_{2}+(A_{x}^{+}+L_{y})n^{x}_{1}}
\end{equation}

\begin{equation}
C_{x,up}^{-}=2\varepsilon_{0}L_{y}L_{z}\frac{1}{F_{x}^{-}-B_{x}^{-}n^{x}_{2}-(A_{x}^{-}+L_{y})n^{x}_{1}}
\end{equation}

\begin{equation}
C_{x,down}^{+}=2\varepsilon_{0}L_{y}L_{z}\frac{1}{F_{x}^{+}+B_{x}^{+}n^{x}_{2}+(A_{x}^{+}-L_{y})n^{x}_{1}}
\end{equation}

\begin{equation}
C_{x,down}^{-}=2\varepsilon_{0}L_{y}L_{z}\frac{1}{F_{x}^{-}-B_{x}^{-}n^{x}_{2}-(A_{x}^{-}-L_{y})n^{x}_{1}}
\end{equation}

To work out the force and the force disturbance along the
sensitive axis we also need to work out capacitance gradients,
which are given by

\begin{equation}
\frac{\partial C_{x,up}^{+}}{ \partial
d_{x}}=\frac{2\varepsilon_{0}L_{y}L_{z}}{\left[F_{x}^{+}+B_{x}^{+}n^{x}_{2}+(A_{x}^{+}+L_{y})n^{x}_{1}\right]^{2}}
\end{equation}

\begin{equation}
\frac{\partial C_{x,up}^{-}}{ \partial
d_{x}}=-\frac{2\varepsilon_{0}L_{y}L_{z}}{\left[F_{x}^{-}-B_{x}^{-}n^{x}_{2}-(A_{x}^{-}+L_{y})n^{x}_{1}\right]^{2}}
\end{equation}

\begin{equation}
\frac{\partial C_{x,down}^{+}}{ \partial
d_{x}}=\frac{2\varepsilon_{0}L_{y}L_{z}}{\left[F_{x}^{+}+B_{x}^{+}n^{x}_{2}+(A_{x}^{+}-L_{y})n^{x}_{1}\right]^{2}}
\end{equation}

\begin{equation}
\frac{\partial C_{x,down}^{-}}{ \partial
d_{x}}=-\frac{2\varepsilon_{0}L_{y}L_{z}}{\left[F_{x}^{-}-B_{x}^{-}n^{x}_{2}-(A_{x}^{-}-L_{y})n^{x}_{1}\right]^{2}}
\end{equation}

Along the y-axis we have

\begin{equation}
\frac{\partial C_{y,up}^{+}}{ \partial
d_{x}}=-\frac{2\varepsilon_{0}L_{x}L_{z}(\phi
+\psi)}{\left[F_{y}^{+}+A_{y}^{+}n^{y}_{1}+(B_{y}^{+}+L_{z})n^{y}_{2}\right]^{2}}
\end{equation}

\begin{equation}
\frac{\partial C_{y,up}^{-}}{ \partial
d_{x}}=\frac{2\varepsilon_{0}L_{x}L_{z}(\phi
+\psi)}{\left[F_{y}^{-}-A_{y}^{-}n^{y}_{1}-(B_{y}^{-}+L_{z})n^{y}_{2}\right]^{2}}
\end{equation}

\begin{equation}
\frac{\partial C_{y,down}^{+}}{ \partial
d_{x}}=-\frac{2\varepsilon_{0}L_{x}L_{z}(\phi
+\psi)}{\left[F_{y}^{+}+A_{y}^{+}n^{y}_{1}+(B_{y}^{+}-L_{z})n^{y}_{2}\right]^{2}}
\end{equation}

\begin{equation}
\frac{\partial C_{y,down}^{-}}{ \partial
d_{x}}=\frac{2\varepsilon_{0}L_{x}L_{z}(\phi
+\psi)}{\left[F_{y}^{-}-A_{y}^{-}n^{y}_{1}-(B_{y}^{-}-L_{z})n^{y}_{2}\right]^{2}}
\end{equation}

Along the z-axis,

\begin{equation}
\frac{\partial C_{z,left}^{+}}{ \partial
d_{x}}=\frac{2\varepsilon_{0}L_{x}L_{y}\phi\theta}{\left[F_{z}^{+}+B_{z}^{+}n^{z}_{2}+(A_{z}^{+}+L_{x})n^{z}_{1}\right]^{2}}
\end{equation}

\begin{equation}
\frac{\partial C_{z,left}^{-}}{ \partial
d_{x}}=-\frac{2\varepsilon_{0}L_{x}L_{y}\phi\theta}{\left[F_{z}^{-}-B_{z}^{-}n^{z}_{2}-(A_{z}^{-}+L_{x})n^{z}_{1}\right]^{2}}
\end{equation}

\begin{equation}
\frac{\partial C_{z,right}^{+}}{ \partial
d_{x}}=\frac{2\varepsilon_{0}L_{x}L_{y}\phi\theta}{\left[F_{z}^{+}+B_{z}^{+}n^{z}_{2}+(A_{z}^{+}-L_{x})n^{z}_{1}\right]^{2}}
\end{equation}

\begin{equation}
\frac{\partial C_{z,right}^{-}}{ \partial
d_{x}}=-\frac{2\varepsilon_{0}L_{x}L_{y}\phi\theta}{\left[F_{z}^{-}-B_{z}^{-}n^{z}_{2}-(A_{z}^{-}-L_{x})n^{z}_{1}\right]^{2}}
\end{equation}

The terms useful for stiffness calculations are variations of
capacitances and capacitance gradients. These are given by

\begin{equation}
\delta
C_{x,up}^{+}=\frac{-2\varepsilon_{0}L_{y}L_{z}}{\left[F_{x}^{+}+B_{x}^{+}n^{x}_{2}+(A_{x}^{+}+L_{y})n^{x}_{1}\right]^{2}}\delta
\left[F_{x}^{+}+B_{x}^{+}n^{x}_{2}+(A_{x}^{+}+L_{y})n^{x}_{1}\right]
\end{equation}

\begin{equation}
\delta
C_{x,up}^{-}=\frac{-2\varepsilon_{0}L_{y}L_{z}}{\left[F_{x}^{-}-B_{x}^{-}n^{x}_{2}-(A_{x}^{-}+L_{y})n^{x}_{1}\right]^{2}}\delta\left[F_{x}^{-}-B_{x}^{-}n^{x}_{2}-(A_{x}^{-}+L_{y})n^{x}_{1}\right]
\end{equation}

\begin{equation}
\delta
C_{x,down}^{+}=\frac{-2\varepsilon_{0}L_{y}L_{z}}{\left[F_{x}^{+}+B_{x}^{+}n^{x}_{2}+(A_{x}^{+}-L_{y})n^{x}_{1}\right]^{2}}\delta\left[F_{x}^{+}+B_{x}^{+}n^{x}_{2}+(A_{x}^{+}-L_{y})n^{x}_{1}\right]
\end{equation}

\begin{equation}
\delta
C_{x,down}^{-}=\frac{-2\varepsilon_{0}L_{y}L_{z}}{\left[F_{x}^{-}-B_{x}^{-}n^{x}_{2}-(A_{x}^{-}-L_{y})n^{x}_{1}\right]^{2}}\delta
\left[F_{x}^{-}-B_{x}^{-}n^{x}_{2}-(A_{x}^{-}-L_{y})n^{x}_{1}\right]
\end{equation}

Finally the last useful formulae for stiffness calculations are
those of the type $\delta C_{i}'$. On the x-axis we have

\begin{equation}
\delta
C_{x,up}^{+'}=\frac{-4\varepsilon_{0}L_{y}L_{z}}{\left[F_{x}^{+}+B_{x}^{+}n^{x}_{2}+(A_{x}^{+}+L_{y})n^{x}_{1}\right]^{3}}\delta
\left[F_{x}^{+}+B_{x}^{+}n^{x}_{2}+(A_{x}^{+}+L_{y})n^{x}_{1}\right]
\end{equation}

\begin{equation}
\delta
C_{x,up}^{-'}=\frac{4\varepsilon_{0}L_{y}L_{z}}{\left[F_{x}^{-}-B_{x}^{-}n^{x}_{2}-(A_{x}^{-}+L_{y})n^{x}_{1}\right]^{3}}
\delta\left[F_{x}^{-}-B_{x}^{-}n^{x}_{2}-(A_{x}^{-}+L_{y})n^{x}_{1}\right]
\end{equation}

\begin{equation}
\delta
C_{x,down}^{+'}=\frac{-4\varepsilon_{0}L_{y}L_{z}}{\left[F_{x}^{+}+B_{x}^{+}n^{x}_{2}+(A_{x}^{+}-L_{y})n^{x}_{1}\right]^{3}}
\delta\left[F_{x}^{+}+B_{x}^{+}n^{x}_{2}+(A_{x}^{+}-L_{y})n^{x}_{1}\right]
\end{equation}

\begin{equation}
\delta
C_{x,down}^{-'}=\frac{4\varepsilon_{0}L_{y}L_{z}}{\left[F_{x}^{-}-B_{x}^{-}n^{x}_{2}-(A_{x}^{-}-L_{y})n^{x}_{1}\right]^{3}}\delta
\left[F_{x}^{-}-B_{x}^{-}n^{x}_{2}-(A_{x}^{-}-L_{y})n^{x}_{1}\right]
\end{equation}

On the y-axis the expressions are as follows:

\begin{equation}
\delta C_{y,up}^{+'}= \frac{4\varepsilon_{0}L_{x}L_{z}(\phi
+\psi)}{\left[F_{y}^{+}+A_{y}^{+}n^{y}_{1}+(B_{y}^{+}+L_{z})n^{y}_{2}\right]^{3}}
\delta\left[F_{y}^{+}+A_{y}^{+}n^{y}_{1}+(B_{y}^{+}+L_{z})n^{y}_{2}\right]
\end{equation}

\begin{equation}
\delta C_{y,up}^{-'}= -\frac{4\varepsilon_{0}L_{x}L_{z}(\phi
+\psi)}{\left[F_{y}^{-}-A_{y}^{-}n^{y}_{1}-(B_{y}^{-}+L_{z})n^{y}_{2}\right]^{3}}
\delta\left[F_{y}^{-}-A_{y}^{-}n^{y}_{1}-(B_{y}^{-}+L_{z})n^{y}_{2}\right]
\end{equation}

\begin{equation}
\delta C_{y,down}^{+'}= \frac{4\varepsilon_{0}L_{x}L_{z}(\phi
+\psi)}{\left[F_{y}^{+}+A_{y}^{+}n^{y}_{1}+(B_{y}^{+}-L_{z})n^{y}_{2}\right]^{3}}
\delta
\left[F_{y}^{+}+A_{y}^{+}n^{y}_{1}+(B_{y}^{+}-L_{z})n^{y}_{2}\right]
\end{equation}

\begin{equation}
\delta C_{y,right}^{-'}= -\frac{4\varepsilon_{0}L_{x}L_{z}(\phi
+\psi)}{\left[F_{y}^{-}-A_{y}^{-}n^{y}_{1}-(B_{y}^{-}-L_{z})n^{y}_{2}\right]^{3}}
\delta\left[F_{y}^{-}-A_{y}^{-}n^{y}_{1}-(B_{y}^{-}-L_{z})n^{y}_{2}\right]
\end{equation}

And finally on the z-axis we have

\begin{equation}
\delta
C_{z,left}^{+'}=-\frac{4\varepsilon_{0}L_{x}L_{y}\phi\theta}{\left[F_{z}^{+}+B_{z}^{+}n^{z}_{2}+(A_{z}^{+}+L_{x})n^{z}_{1}\right]^{3}}
\delta\left[F_{z}^{+}+B_{z}^{+}n^{z}_{2}+(A_{z}^{+}+L_{x})n^{z}_{1}\right]
\end{equation}

\begin{equation}
\delta
C_{z,left}^{-'}=\frac{4\varepsilon_{0}L_{x}L_{y}\phi\theta}{\left[F_{z}^{-}-B_{z}^{-}n^{z}_{2}-(A_{z}^{-}+L_{x})n^{z}_{1}\right]^{3}}
\delta\left[F_{z}^{-}-B_{z}^{-}n^{z}_{2}-(A_{z}^{-}+L_{x})n^{z}_{1}\right]
\end{equation}

\begin{equation}
\delta
C_{z,right}^{+'}=-\frac{4\varepsilon_{0}L_{x}L_{y}\phi\theta}{\left[F_{z}^{+}+B_{z}^{+}n^{z}_{2}+(A_{z}^{+}-L_{x})n^{z}_{1}\right]^{3}}
\delta\left[F_{z}^{+}+B_{z}^{+}n^{z}_{2}+(A_{z}^{+}-L_{x})n^{z}_{1}\right]
\end{equation}

\begin{equation}
\delta
C_{z,right}^{-'}=\frac{4\varepsilon_{0}L_{x}L_{y}\phi\theta}{\left[F_{z}^{-}-B_{z}^{-}n^{z}_{2}-(A_{z}^{-}-L_{x})n^{z}_{1}\right]^{3}}
\delta\left[F_{z}^{-}-B_{z}^{-}n^{z}_{2}-(A_{z}^{-}-L_{x})n^{z}_{1}\right]
\end{equation}
where for the x, y and z axes we have

\begin{equation}
\delta\left[F_{x}^{+}+B_{x}^{+}n^{x}_{2}+(A_{x}^{+}\pm
L_{y})n^{x}_{1}\right]=-\delta d_{x}\pm L_{y}(\delta
\phi+\delta\psi)
\end{equation}

\begin{equation}
\delta\left[F_{x}^{-}-B_{x}^{-}n^{x}_{2}-(A_{x}^{-}\pm
L_{y})n^{x}_{1}\right]=+\delta d_{x}\mp L_{y}(\delta
\phi+\delta\psi)
\end{equation}

\begin{equation}
\delta\left[F_{y}^{+}+A_{y}^{+}n^{y}_{1}+(B_{y}^{+}\pm
L_{z})n^{y}_{2}\right]=-\delta d_{y}\pm L_{z}\delta \theta
\end{equation}

\begin{equation}
\delta\left[F_{y}^{-}-A_{y}^{-}n^{y}_{1}-(B_{y}^{-}\pm
L_{z})n^{y}_{2}\right]=\delta d_{y}\mp L_{z}\delta \theta
\end{equation}

\begin{equation}
\delta\left[F_{z}^{+}+B_{z}^{+}n^{z}_{2}+(A_{z}^{+}\pm
L_{x})n^{z}_{1}\right]=-\delta d_{z}
\end{equation}

\begin{equation}
\delta\left[F_{z}^{-}-B_{z}^{-}n^{z}_{2}-(A_{z}^{-}\pm
L_{x})n^{z}_{1}\right]=\delta d_{z}
\end{equation}

\subsection{Capacitance as a position sensor.}
By measuring and combining  capacitances along the different axis,
we can obtain the position and attitude of the proof mass. The
parameters $(d_{x},d_{y},d_{z},\phi, \theta, \psi)$ are obtained
by the following combination of capacitances,

\begin{equation}
(C_{z,r}^{+}-C_{z,r}^{-})+(C_{z,l}^{+}-C_{z,l}^{-})\simeq8\varepsilon_{0}L_{x}L_{y}\frac{d_{z}}{D_{z}^{2}}
\end{equation}

\begin{equation}
(C_{z,r}^{+}-C_{z,r}^{-})-(C_{z,l}^{+}-C_{z,l}^{-})\simeq8\varepsilon_{0}\frac{L_{x}^{2}L_{y}}{D_{z}^{2}}\theta\phi
\end{equation}

\begin{equation}
(C_{x,up}^{+}-C_{x,up}^{-})+(C_{x,down}^{+}-C_{x,down}^{-})\simeq8\varepsilon_{0}L_{z}L_{y}\frac{d_{x}}{D_{x}^{2}}
\end{equation}

\begin{equation}
(C_{x,up}^{+}-C_{x,up}^{-})-(C_{x,down}^{+}-C_{x,down}^{-})\simeq-8\varepsilon_{0}L_{y}^{2}L_{z}(\phi+\psi)\frac{1}{D_{x}^{2}}
\end{equation}

\begin{equation}
(C_{y,up}^{+}-C_{y,up}^{-})+(C_{y,down}^{+}-C_{y,down}^{-})\simeq8\varepsilon_{0}L_{x}L_{z}\frac{d_{y}}{D_{y}^{2}}
\end{equation}

\begin{equation}
(C_{y,up}^{+}-C_{y,up}^{-})-(C_{y,down}^{+}-C_{y,down}^{-})\simeq-8\varepsilon_{0}L_{x}L_{z}^{2}\theta\frac{1}{D_{y}^{2}}
\end{equation}

\subsection{Capacitances, capacitance derivatives and their variations}

For the special case in which the proof mass is in the equilibrium
position, $\vec{0}$, with no translational and rotational offsets,

\begin{equation}
C_{x,up}^{\pm}=C_{x,down}^{\pm}=\frac{2\varepsilon_{0}L_{y}L_{z}}{D_{x}}
\end{equation}

\begin{equation}
C_{y,up}^{\pm}=C_{y,down}^{\pm}=\frac{2\varepsilon_{0}L_{x}L_{z}}{D_{y}}
\end{equation}

\begin{equation}
C_{z,left}^{\pm}=C_{z,right}^{\pm}=\frac{2\varepsilon_{0}L_{x}L_{y}}{D_{z}}
\end{equation}

\begin{equation}
\frac{\partial C_{x,up}^{\pm}}{\partial d_{x}}=\frac{\partial
C_{x,down}^{\pm}}{\partial
d_{x}}=\pm\frac{2\varepsilon_{0}L_{y}L_{z}}{D_{x}^{2}}
\end{equation}

\begin{equation}
\frac{\partial C_{y,up}^{\pm}}{\partial d_{x}}=\frac{\partial
C_{y,down}^{\pm}}{\partial d_{x}}=\frac{\partial
C_{z,left}^{\pm}}{\partial d_{x}}=\frac{\partial
C_{z,right}^{\pm}}{\partial d_{x}}\approx0
\end{equation}

\begin{equation}
\delta
C_{x,up}^{\pm}=\mp\frac{2\varepsilon_{0}L_{y}L_{z}}{D_{x}^{2}}\left[-\delta
d_{x}+ L_{y}(\delta \phi+\delta\psi)\right]
\end{equation}

\begin{equation}
\delta
C_{x,down}^{\pm}=\mp\frac{2\varepsilon_{0}L_{y}L_{z}}{D_{x}^{2}}\left[-\delta
d_{x}- L_{y}(\delta \phi+\delta\psi)\right]
\end{equation}

\begin{equation}
\delta
C_{y,up}^{\pm}=\mp\frac{2\varepsilon_{0}L_{x}L_{z}}{D_{y}^{2}}\left[-\delta
d_{y}+L_{z}\delta \theta\right]
\end{equation}

\begin{equation}
\delta
C_{y,down}^{\pm}=\mp\frac{2\varepsilon_{0}L_{x}L_{z}}{D_{y}^{2}}\left[-\delta
d_{y}-L_{z}\delta\theta\right]
\end{equation}

\begin{equation}
\delta C_{z,left}^{\pm}=\delta
C_{z,right}^{\pm}=\pm\frac{2\varepsilon_{0}L_{x}L_{y}}{D_{z}^{2}}\delta
d_{z}
\end{equation}

\begin{equation}
\delta
C_{x,up}^{\pm'}=\frac{-4\varepsilon_{0}L_{y}L_{z}}{D_{x}^{3}}\left[-\delta
d_{x}+ L_{y}(\delta \phi+\delta\psi)\right]
\end{equation}

\begin{equation}
\delta
C_{x,down}^{\pm'}=\frac{-4\varepsilon_{0}L_{y}L_{z}}{D_{x}^{3}}\left[-\delta
d_{x}- L_{y}(\delta \phi+\delta\psi)\right]
\end{equation}

\begin{equation}
\delta C_{y,up}^{\pm'}=\delta C_{y,down}^{\pm'}=\delta
C_{z,left}^{\pm'}=\delta C_{z,right}^{\pm'}\approx0
\end{equation}

\end{document}